# Spin injection into silicon detected by broadband ferromagnetic resonance spectroscopy

Ryo Ohshima,[1] Stefan Klingler,[2,3], Sergey Dushenko,[1] Yuichiro Ando,[1] Mathias Weiler,[2,3] Hans Huebl,[2,3,4] Teruya Shinjo,[1] Sebastian T. B. Goennenwein,[2,3,4] and Masashi Shiraishi[1*]

[1]Department of Electronic Science and Engineering, Kyoto Univ., 615-8510 Kyoto, Japan.
[2]Walther-Meißner-Institut, Bayerische Akademie der Wissenschaften, 85748 Garching, Germany.
[3]Physik-Department, Technische Universität München, 85748 Garching, Germany
[4]Nanosystems Initiative Munich, 80799 München, Germany

**We studied the spin injection in a NiFe(Py)/Si system using broadband ferromagnetic resonance spectroscopy. The Gilbert damping parameter of the Py layer on top of the Si channel was determined as a function of the Si doping concentration and Py layer thickness. For fixed Py thickness we observed an increase of the Gilbert damping parameter with decreasing resistivity of the Si channel. For a fixed Si doping concentration we measured an increasing Gilbert damping parameter for decreasing Py layer thickness. No increase of the Gilbert damping parameter was found Py/Si samples with an insulating interlayer. We attribute our observations to an enhanced spin injection into the low-resistivity Si by spin pumping.**

Spin injection into semiconductors was relentlessly studied in recent years in hope to harness their long spin relaxation time, and gate tunability to realize spin metal-oxide-semiconductor field-effect-transistors (MOSFETs). A central obstacle for a spin injection into semiconductors was the conductance mismatch[1] between ferromagnetic metals (used for the spin injection) and semiconductor channels. In an electrical spin injection method—widely used from the early years of the non-local spin transport studies—tunnel barriers between the semiconductor and ferromagnet were formed to avoid the conductance mismatch problem[2–5]. Unfortunately, it complicated the production process of the devices, as high quality tunnel barriers are not easy to grow, and presence of impurities, defects and pinholes takes a heavy toll on the spin injection efficiency and/or induces spurious effects. Meanwhile, the fabrication of electrical Si spin devices, like spin MOSFETs, with different resistivities is a time-consuming process, which so far prevented systematic studies of the spin injection properties (such as spin lifetime, spin injection efficiency etc.). However, such a systematic study is necessary for further progress towards practical applications of spin MOSFETs.

In 2002, a dynamical spin injection method, known as spin pumping, was introduced to the scene of spintronics research[6,7]. While the method was initially used in the metallic multilayer systems, it was later implemented to inject spin currents into semiconductors. In contrast to electrical spin injection, spin pumping does not require the application of an electric current across the ferromagnet/semiconductor interface. Devices that operate using spin currents instead of charge currents can potentially reduce heat generation and power consumption problems of modern electronics. From a technological point of view, spin pumping is also appealing because it does not require a tunnel barrier. Spin injection—using spin pumping—into semiconductors from an adjacent ferromagnetic metal was achieved despite the existence of conductivity mismatch[8-12]. However, so far there was no systematic study of the spin pumping based spin injection in dependence on the resistivity of the Si channel.

In this letter, we focus on the study of spin injection by spin pumping in the NiFe(Py)/Si system with different resistivities of the Si channel using broadband ferromagnetic resonance (FMR). The broadband FMR method allows for a precise determination of the Gilbert damping parameter $\alpha$, which increases in the presence of spin pumping, and thus, spin injection. By tracking the change of the Gilbert damping parameter in various Py/Si systems, we determined the spin pumping efficiency in the broad range of resistivities of the Si channel.

For a first set of samples 7nm-thick Py films were deposited by electron beam evaporation on top of various Si substrates ($1\times1$ $cm^2$ in size) with resistivities in the range from $10^{-3}$ to $10^3$ $\Omega\cdot$cm (see Table 1 for the list of the prepared samples). The oxidized surface of the Si substrates was removed using 10% hydrofluoric acid (HF) prior to the Py evaporation. For a second set of samples Py films with thicknesses $d_{\mathrm{Py}}$ between 5nm and 80nm were deposited on P-doped SOI (silicon on insulator) with the same technique. As a control experiment, Py/$AlO_x$ and Py/$TiO_x$ films were grown on Si, P-doped SOI and $SiO_2$ substrates, as spin pumping should be suppressed in systems with an insulating barrier[13] (see Table 2). Both Al (3 nm, thermal deposition) and Ti (2 nm, electron beam evaporation) were evaporated on the non-treated substrates and left in the air for one day for oxidation of the surface (for the Al layer, the process was repeated 3 times, with 1 nm of Al evaporated and oxidized at each step). After oxidation, we evaporated 7nm thick Py films on the top of the tunnel barriers. The properties of the prepared samples are summarized in Tables 1 and 2.

A sketch of the broadband ferromagnetic resonance setup is shown in Fig. 1(a). The samples were placed face down on the center conductor of a coplanar waveguide (CPW), which was located between the pole shoes of an electromagnet. A static magnetic field $|\mu_0 H| \leq 2.5$ T was applied perpendicular to the surface of the samples to avoid extra damping due to two-magnon scattering[14]. One end of the CPW was connected to a microwave source, where microwaves with frequency $f < 40$ GHz were generated. The other end of the CPW was connected to a microwave diode and a lock-in amplifier to measure the rectified microwave voltage as a function of the applied magnetic field. All measurements were carried out at room temperature.

The microwave current in the CPW generates an oscillating magnetic field around the center conductor which results in an oscillating torque on the sample's magnetization. For $\mu_0 H = \mu_0 H_{\mathrm{FMR}}$ this torque results in an absorption of microwave power. The resonance condition is given by the out-of-plane Kittel equation[15,16]:

$$\frac{hf}{g\mu_{\mathrm{B}}} = \mu_0 H_{\mathrm{FMR}} - \mu_0 M_{\mathrm{eff}}. \tag{1}$$

Here, $h$ is the Planck constant, $g$ is the Landé $g$-factor, $\mu_{\mathrm{B}}$ is the Bohr magneton, $\mu_0$ is the vacuum permeability, and $M_{\mathrm{eff}}$ is the effective saturation magnetization of Py.

We use the Gilbert damping model, which phenomenologically models the viscous damping of the magnetic resonance. The linear relation between the full width at half maximum $\Delta H$ of the resonance and the applied microwave frequency $f$ is given by the Gilbert damping equation [17]:

$$\mu_0 \Delta H = \mu_0 \Delta H_0 + \frac{2\alpha hf}{g\mu_{\mathrm{B}}}. \tag{2}$$

Here, $\Delta H_0$ corresponds to frequency independent scattering processes and $\alpha$ is the Gilbert damping parameter[18,19]:

$$\alpha = \alpha_0 + \alpha_{\mathrm{SP}} + \alpha_{\mathrm{EC}}. \tag{3}$$

Here, $\alpha_0$ is the intrinsic Gilbert damping, $\alpha_{\mathrm{SP}} = g\mu_B g_{\mathrm{r}}^{\uparrow\downarrow}/4\pi M_{\mathrm{S}} d_{\mathrm{Py}}$ is the damping due to spin pumping[20], $g_{\mathrm{r}}^{\uparrow\downarrow}$ is the real part of the spin mixing conductance, and $\alpha_{\mathrm{EC}} = C_{\mathrm{EC}} d_{\mathrm{Py}}^2$ is the eddy-current damping. The parameter $C_{\mathrm{EC}}$ describes efficiency of the eddy-current damping.

To realize a net spin injection via spin pumping, the following conditions should be fulfilled in the system: (i) carriers should be present in the underlying channel, (ii) the spin relaxation time in the channel should be small enough. The available carriers in the channel transfer spin angular momentum away from the spin injection interface, allowing propagation of the spin current. On the other hand, the long spin relaxation time in the channel leads to a large spin accumulation at the interface and generates a diffusive spin backflow in the direction opposite to the spin pumping current[20] (see Figs. 1(b) and 1(c)). Thus, the spin backflow effectively cancels out the spin pumping current for long spin relaxation time, and the spin pumping contribution to the Gilbert damping parameter should no longer be present in the system.

In addition to spin pumping, charge currents can be induced in the Si channel and Py layer due to the Faraday's law and Py magnetization precession, which results in a Gilbert-like damping contribution. These processes are referred to as radiative damping and eddy-current damping[18]. An enhancement of the Gilbert damping parameter due to these processes is expected to be especially large for the Si channels with low resistivities and thick Py films, since energy dissipation through eddy currents scales linearly with the conductivity of the Si layer and quadratically with the Py layer thickness. Hence, both spin pumping and eddy current damping are expected to be most efficient for low-resistivity Si. In contrast to spin pumping, the radiative damping contribution does not require a direct electrical contact between Py and Si, and is hence unaffected by the tunnel barrier.

Figure 1(d) shows a typical FMR spectrum of a 7nm thick Py film on a phosphorous P-doped Si on insulator (SOI) substrate, where the microwave frequency was fixed at 30 GHz during the sweep of the magnetic field. A single FMR signal was observed (Fig. 1(d) red filled circles), from which $\mu_0 H_{\mathrm{FMR}}$ and $\mu_0 \Delta H$ were extracted by a fit of the magnetic ac susceptibility (Fig. 1(d) black line) [16,21] (see Supplemental Material for additional fitting examples). An excellent agreement of the fit with the measurement is achieved.

Figures 2(a) and (b) show $H_{\mathrm{FMR}}$ and $\Delta H$ versus the applied microwave frequency $f$ for the Py/P-doped SOI, Py/SOI and Py/$SiO_2$ samples. From the fitting of the frequency dependence of $H_{\mathrm{FMR}}$ with Eq.(1), $g$ and $\mu_0 M_{\mathrm{eff}}$ of the Py/P-doped SOI (Py/SOI) were estimated to be 2.049 (2.051) and 0.732 T (0.724 T), and those of the Py/$SiO_2$ were estimated to be 2.038 and 0.935 T, respectively (Supplemental Material for fitting and data from other samples). The difference in the $g$ and $M_{\mathrm{eff}}$ between the Py/Si and the Py/$SiO_2$ sample is attributed to the inter-diffusion of the Fe/Ni and Si at the interface, which is always present to some extent during the growth at room temperature[22,23]. The Gilbert damping $\alpha$ of the Py/P-doped SOI, Py/SOI and Py/$SiO_2$ were estimated to be $1.25\times10^{-2}$, $9.02\times10^{-3}$ and $8.49\times10^{-3}$, respectively, from the linewidth vs. frequency evolution. The intrinsic Gilbert damping parameter $\alpha_0$ is determined from the linewidth evolution of the Py/$SiO_2$ and Py/quartz samples to be $8.5\times10^{-3}$ and $8.6\times10^{-3}$, respectively, since no spin pumping contribution is expected in these insulating materials ($\alpha = \alpha_0$). From this we can see an increasing Gilbert damping with decreasing resistivity. We additionally measured the samples with an insulating tunnel barrier between the Si channel and the Py film. We found Gilbert damping parameters of $\alpha = 8.8\times10^{-3}$ for the Py/$AlO_x$/Si samples and $\alpha = 7.5\times10^{-3}$ for the Py/$TiO_x$/Si samples, independent of the Si resistivity. The damping values are in agreement with the intrinsic damping extracted from the Py/$SiO_2$ sample, indicating that radiative damping is negligible in our samples.

Figure 3(a) summarizes the dependence of the Gilbert damping parameter $\alpha$ on the resistivity of the Si channel (see Supplemental Material D for the g-factor, the effective saturation magnetization and the frequency independent term), including the measured control samples. The dashed lines show the intrinsic contributions $\alpha_0$ to the Gilbert damping parameter $\alpha$ measured from the Py/$SiO_2$ and Py/quartz samples (red dashed), Py/$AlO_x$/$SiO_2$ (blue dashed) and Py/$TiO_x$/$SiO_2$ (green dashed). All samples with Py on top of the conductive substrates without an additional tunnel barrier exhibited the Gilbert damping parameter $\alpha$ larger than the intrinsic contribution $\alpha_0$.

The experimentally measured Gilbert damping parameter decreases logarithmically with the resistivity. This result is in agreement with condition (i) for the spin pumping. In the Si channels with a small resistivity more carriers were available to transfer the injected angular momentum, leading to an effective spin pumping. Additionally, both electron spin resonance[25–28] and non-local 4-terminal Hanle precession[29,30] experiments showed, that the spin lifetime in Si is increasing with increasing resistivity. Our samples with low resistivities have a large doping concentration (see Table 1), leading to shorter spin relaxation time. In accordance with the spin pumping condition (ii), the decrease of the spin relaxation time should lead to the increase of the spin pumping contribution, as now observed experimentally. While the spin pumping shows a logarithmic dependence on the resistivity of the channel, we note that an increased Gilbert damping parameter  is observed even for the Si channel with high resistivities. We comment on the Sb-doped sample, where the experimentally measured $\alpha_{\mathrm{SP}}$ was lower than one expected from the logarithmic trend of the other samples. We speculate that this might originate from the different doping profile, compared to the other samples. We note, that further studies are necessary to separate the influence of the number of carriers in the channel and the spin relaxation time on the spin pumping process.

Finally, we show that the Py damping is increased in a broad range of Si resistivities and attribute this effect to the enhanced spin injection via spin pumping (a discussion of the spin mixing conductance for various Si resistivities is given in the Supplemental Material A). Figure 3(b) shows the Py thickness dependence of the $\alpha$ for Py/P-doped SOI samples. The solid line shows a fit of Eq.(3) to the measured data and a very good agreement of the spin pumping theory with our measurements is achieved. From the fit we estimate $\alpha_0 = 6.1\times10^{-3}$, $g_{\mathrm{r}}^{\uparrow\downarrow} = 1.2\times10^{19}$ m$^{-2}$ and $C_{\mathrm{EC}} = 2.9\times10^{11}$ m$^{-2}$. Both the intrinsic damping and the real part of the spin mixing conductance are in good agreement with previous measurements[30]. The blue dashed line indicates $\alpha_0 + \alpha_{\mathrm{SP}}$ and the green dashed line shows $\alpha_0 + \alpha_{\mathrm{EC}}$. Dominant influence of spin pumping to the total damping is observed in samples with small Py thickness, while eddy current contribution is dominant in samples with thick Py layer. For the 7 nm-thick Py sample, we find $\alpha_{\mathrm{SP}}$ = 3.6×10$^{-3}$ and $\alpha_{\mathrm{EC}}$ = 1.4×10$^{-5}$. Thus, the eddy-current damping in our 7 nm Py samples is negligibly small and cannot explain the increase of the damping with decreasing resistivity. The Py thickness dependence of the Gilbert damping indicates spin pumping into the Si substrates.

In conclusion, we studied spin pumping based spin injection from a Py layer into Si channels with various resistivities using broadband ferromagnetic resonance. We determined the spin pumping contribution from the change of the Gilbert damping parameter. The observed logarithmic decrease of the Gilbert damping parameter with increasing resistivity of the Si channel is attribute to the decrease in the number of carriers in the channel, and the increase in the spin lifetime. Despite the reduction of the spin pumping contribution to the Gilbert damping parameter with the increasing resistivity of the Si channel, we observe spin pumping even for the channels with high resistivity. We furthermore observe an increase of the Gilbert damping parameter for decreasing Py thickness which is in agreement with the spin pumping theory. Our results show that spin pumping can be potentially used in a spin transistors, where low doping concentration in the channel is necessary for the gate control of the device.

**Supplemental Material**

See Supplementary Material for a discussion of the spin mixing conductance for various Si resistivities and additional fitting examples.

**ACKNOWLEGEMENTS**

This research was supported in part by a Grant-in-Aid for Scientific Research from the Ministry of Education, Culture, Sports, Science and Technology (MEXT) of Japan, Innovative Area “Nano Spin Conversion Science” (No. 26103003), Scientific Research (S)

"Semiconductor Spincurrentronics" (No. 16H0633) and JSPS KAKENHI Grant (No. 16J00485). R.O. acknowledges JSPS Research Fellowship. S.D. acknowledges support by JSPS Postdoctoral Fellowship and JSPS KAKENHI Grant No. 16F16064.

mixing conductance are due to slightly different growth conditions.

Figure 1: (a) Experimental setup for the broadband FMR measurement. The samples were placed with the Py layer facing down on a coplanar waveguide. External magnetic field and microwave field from the waveguide induce the FMR of the Py and spins are injected into Si via spin pumping. Schematic images of spin injection and dephasing in Si that have (b) long and (c) short spin lifetimes. $\tau_1$ and $\tau_2$ are the spin lifetime of Si in the case of (b) and (c), respectively. Spin injection efficiency becomes large in the case of (c) because of a reduction of the backflow of spins. (d) The derivative of the FMR signal of Py at 30 GHz microwave frequency. $I$ is the microwave absorption intensity.

Figure 2: Frequency dependence of the (a) resonance field $H_{\mathrm{FMR}}$ and (b) full width at half maximum $\Delta H$ of the FMR spectra obtained from Py on top of P-doped SOI, SOI and $SiO_2$. The solid lines show fitting using Eqs. (1) and (2) of $H_{\mathrm{FMR}}$ and $\Delta H$, respectively.

Figure 3: (a) Si resistivity dependence of the Gilbert damping parameter $\alpha$. The damping of the samples with an insulating layer represents the intrinsic damping of the Py layer and is shown by the dashed lines. Red, blue and green coloration represents Py, Py/$AlO_x$ and Py/$TiO_x$ samples, respectively. The damping of the Py/P-doped SOI is an averaged value extracted from the two Py/P-doped SOI samples fabricated at different times. (b) Py thickness dependence of $\alpha$. The solid line shows a fit of Eq. (3) to the data. The blue line shows $\alpha_0 + \alpha_{\mathrm{SP}}$, whereas the green line shows $\alpha_0 + \alpha_{\mathrm{EC}}$.

Fig. 1 R. Ohshima et al.

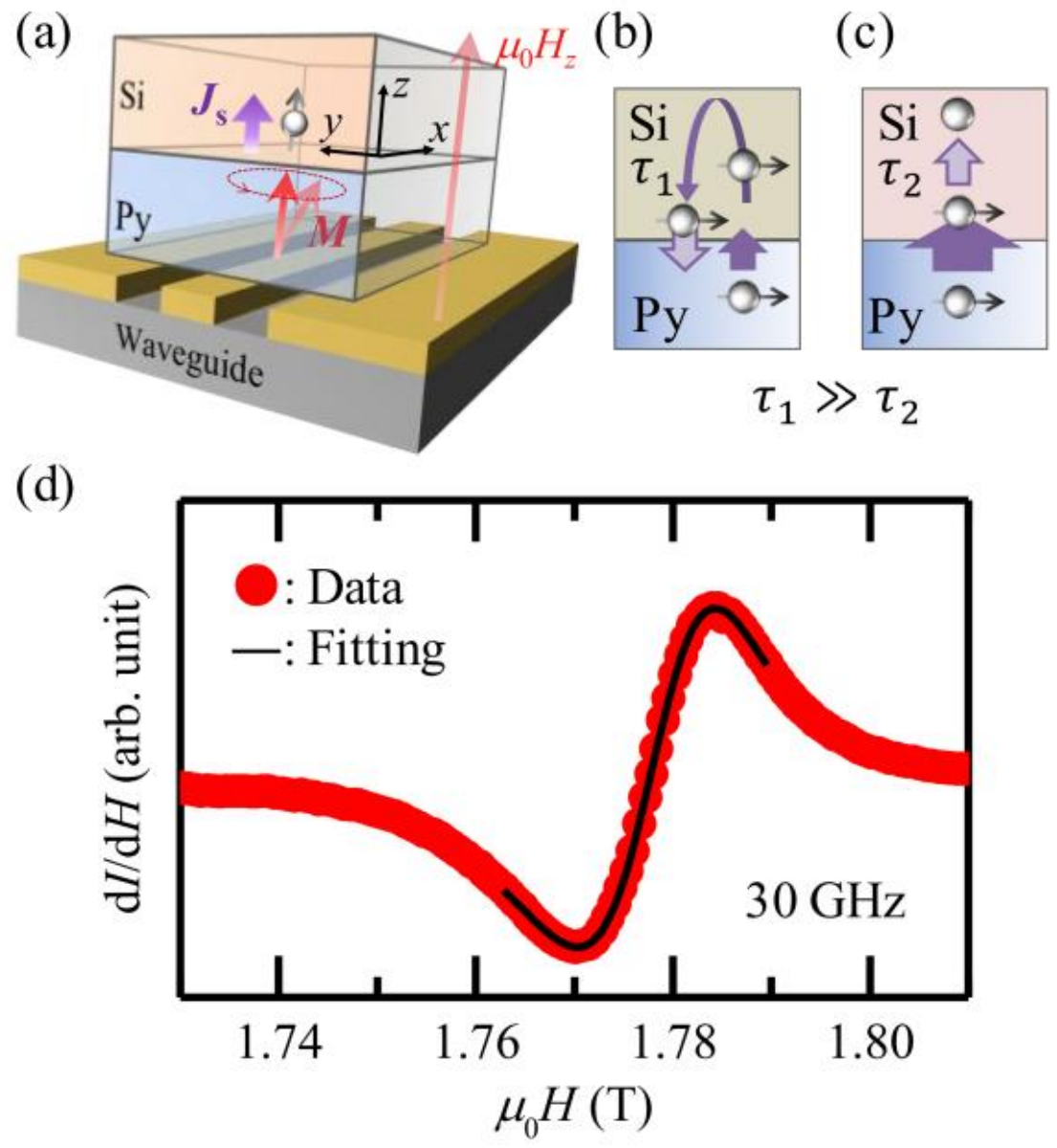

Fig. 2 R. Ohshima et al.

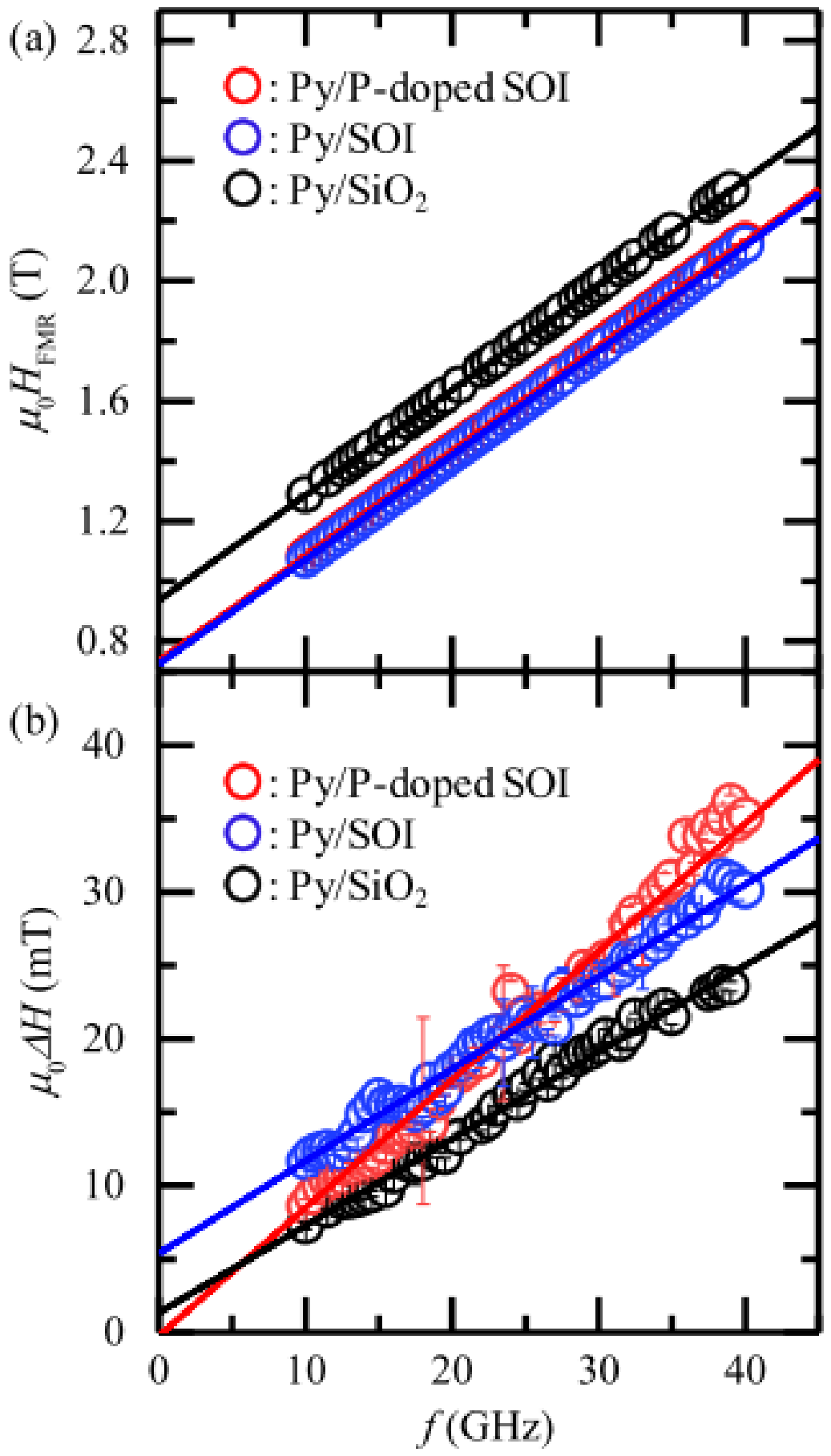

Fig. 3    R. Ohshima et al.

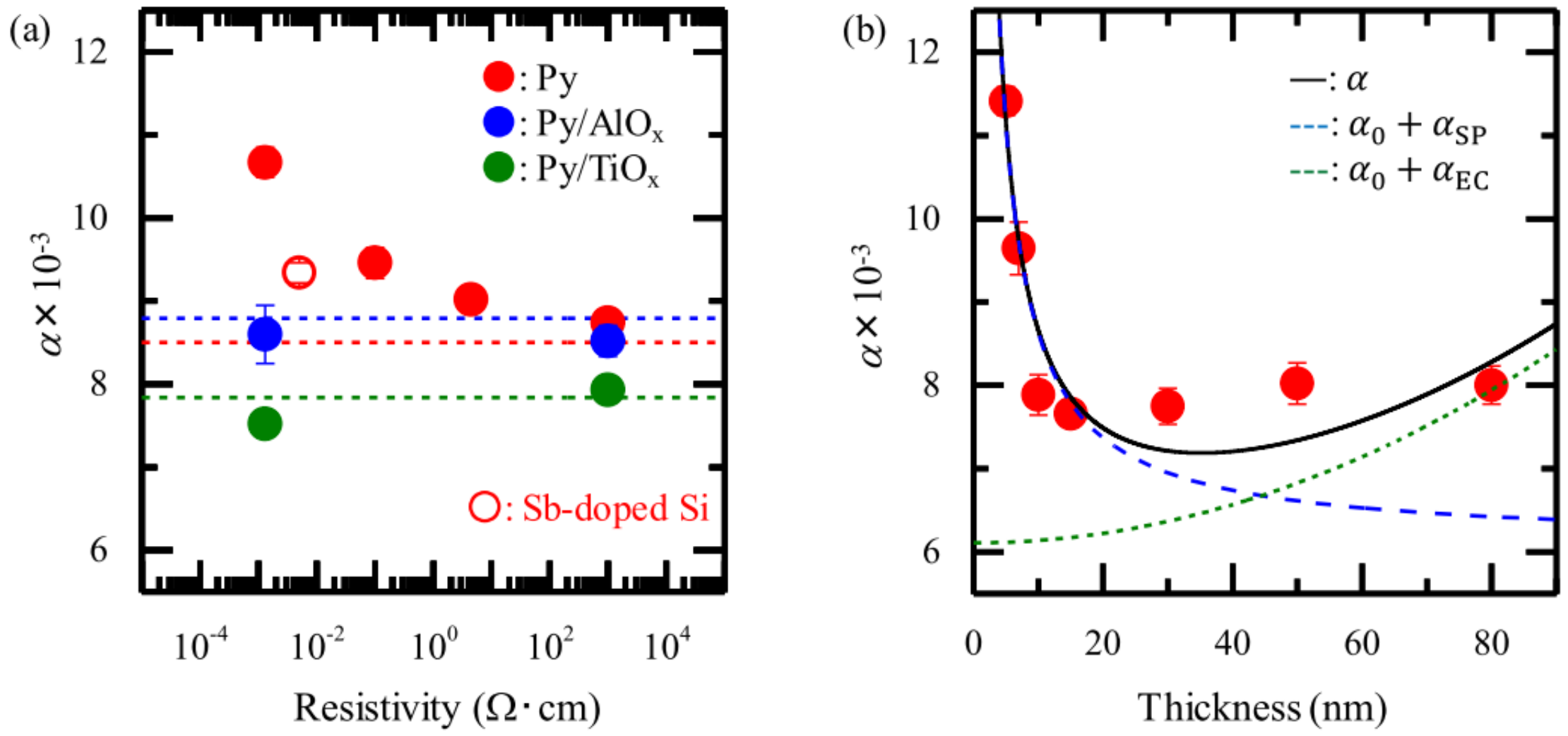

Table 1: Sample summary

| Name | Dopant | Doping density ($cm^{-3}$) | Structure | Resistivity (Ω・cm) | Gilbert damping parameter | Mixing conductance ($m^{-2}$) |
|---|---|---|---|---|---|---|
| Py/P-doped SOI | P | $6.5\times10^{19}$ | Py(7 nm)/Si(100 nm)/ $SiO_2$(200 nm)/Si | $1.3\times10^{-3}$ | $1.1\times10^{-2}$ | $5.7\times10^{18}$ |
| Py/Sb-doped Si | Sb | $1\times10^{19}$ | Py(7 nm)/Si | $5.0\times10^{-3}$ | $9.3\times10^{-3}$ | $2.3\times10^{18}$ |
| Py/N-doped Si | N | $1\times10^{19}$ | Py(7 nm)/Si | $1.0\times10^{-1}$ | $9.5\times10^{-3}$ | $2.6\times10^{18}$ |
| Py/SOI | N/A | $1\times10^{15}$ | Py(7 nm)/Si(100 nm)/ $SiO_2$(200 nm)/Si | 4.5 | $9.0\times10^{-3}$ | $1.3\times10^{18}$ |
| Py/P-doped Si | P | $1\times10^{13}$ | Py(7 nm)/Si | $1.0\times10^{3}$ | $8.7\times10^{-3}$ | $5.1\times10^{17}$ |
| Py/$SiO_2$ | - | - | Py(7 nm) /$SiO_2$(500 nm)/Si | - | $8.5\times10^{-3}$ | - |
| Py/Quartz | - | - | Py(7 nm)/Quartz | - | $8.6\times10^{-3}$ | - |

Table 2: List of the samples for the control experiment

| Name | Dopant | Doping density ($cm^{-3}$) | Structure | Resistivity (Ω・cm) | Gilbert damping parameter |
|---|---|---|---|---|---|
| Py/$AlO_x$/P-doped SOI | P | $6.5\times10^{19}$ | Py(7 nm)/$AlO_x$(3 nm)/Si(100 nm)/ $SiO_2$(200 nm)/Si | $1.3\times10^{-3}$ | $8.6\times10^{-3}$ |
| Py/$AlO_x$/P-doped Si | P | $1\times10^{13}$ | Py(7 nm)/$AlO_x$(3 nm)/Si | $1.0\times10^{3}$ | $8.5\times10^{-3}$ |
| Py/$AlO_x$/ $SiO_2$ | - | - | Py(7 nm)/$AlO_x$(3 nm)/ $SiO_2$(500 nm)/Si | - | $8.8\times10^{-3}$ |
| Py/$TiO_x$/P-doped SOI | P | $6.5\times10^{19}$ | Py(7 nm)/$TiO_x$(2 nm)/Si(100 nm)/ $SiO_2$(200 nm)/Si | $1.3\times10^{-3}$ | $7.5\times10^{-3}$ |
| Py/$TiO_x$/P-doped Si | P | $1\times10^{13}$ | Py(7 nm)/$TiO_x$(2 nm)/Si | $1.0\times10^{3}$ | $7.9\times10^{-3}$ |
| Py/$TiO_x$/ $SiO_2$ | - | - | Py(7 nm)/$TiO_x$(2 nm)/ $SiO_2$(500 nm)/Si | - | $7.8\times10^{-3}$ |